\documentclass[preprint]{elsart4}

\usepackage{graphicx}
\usepackage{amssymb}

\begin{document}
  
  \begin{frontmatter}
    
    \title{Ising and anisotropic Heisenberg magnets with mobile defects}
    
    \author[aff1]{M.~Holtschneider}
    \author[aff1]{R.~Leidl}
    \author[aff1]{W.~Selke\corauthref{cor1}}
    
    \address[aff1]{Institut f\"ur Theoretische Physik, RWTH Aachen, 52056--Aachen, Germany}
    \corauth[cor1]{Tel:~++49-(0)241-80-27029; fax:~++49-(0)241-80-22188}
    \ead{selke@physik.rwth-aachen.de}
    
    \begin{abstract}
      Motivated by experiments on
      (Sr,Ca,La)$_{14}$Cu$_{24}$O$_{41}$, a two--dimensional Ising model
      with mobile defects and a two--dimensional anisotropic
      Heisenberg antiferromagnet have been proposed and studied recently. We extend
      previous investigations by analysing phase diagrams of
      both models in external fields using mainly Monte Carlo techniques. In
      the Ising case, the phase transition is due to the thermal
      instability of defect stripes, in the Heisenberg case additional
      spin--flop structures play an essential role. 
    \end{abstract}
    
  \end{frontmatter}
  
\section{Introduction}\label{sec1}
  
The compounds (Sr,La,Ca)$_{14}$Cu$_{24}$O$_{41}$ display interesting
low--dimensional magnetic features arising from Cu$_2$O$_3$
two--leg ladders and CuO$_2$ chains. Here, we
shall consider two types of models which have been
motivated by experimental observations
on the CuO$_2$ chains \cite{miz,amm,mat,klin}.

On the one hand, a simple two--dimensional Ising model with
mobile defects has been introduced \cite{selke}. The spins
correspond to the magnetic Cu$^{2+}$ ions, and the defects
to those Cu ions which are believed to be
spinless due to holes (Zhang--Rice singlets). The
defects have been shown to form, at low
temperatures, nearly straight stripes, perpendicular
to the CuO$_2$ chains. The coherency of the stripes gets
lost at a phase transition of first order \cite{selke,holtschn}.

On the other hand, Matsuda {\it et al.} \cite{mat} proposed a
two--dimensional anisotropic Heisenberg model, with an easy
spin axis, which
reproduced nicely the measured spin--wave dispersions in
La$_5$Ca$_9$Cu$_{24}$O$_{41}$. In
subsequent analyses of this model \cite{leidl1,leidl2}, the effect of
external fields parallel (leading to a spin--flop phase) and
perpendicular to the easy axis
on various thermal quantities has been analysed and
compared to experimental findings \cite{amm,klin}.

In this contribution, we shall extend the previous work
on both models in external fields. For the Ising model with a local pinning
potential, the phase diagram in the (temperature, field)--plane is
determined at various pinning strengths. In the Heisenberg case, the
boundary of the antiferromagnetic phase and the transition to the
spin--flop phase are studied in detail, both for the model of
Matsuda {\it et al.} as well as for a simpler, more common variant of
the anisotropic Heisenberg antiferromagnet \cite{bl}.

\section{Ising model with mobile defects}\label{sec2}

The Ising model with mobile defects is defined on
a square or rectangular lattice
with one axis defining the chain direction (horizontal direction
in Fig. 1) \cite{selke,holtschn}. Each
lattice site is occupied either by a spin, $S_i= \pm 1$, or
by a defect, $S_i= 0$. Usually, the concentration of
defects is fixed to be ten percent of the lattice sites, as it seems to
be the case in La$_5$Ca$_9$Cu$_{24}$O$_{41}$. Defects are assumed
to be mobile along the
chains, keeping a minimal distance of two lattice
spacings. Neighbouring spins are coupled
ferromagnetically, $J >0$, along the chains, and
antiferromagnetically, $J_a < 0$ perpendicular to
them. In addition, next--nearest neighbouring spins
in the same chain separated by a defect interact
antiferromagnetically, $J_0 <0$, as suggested by experiments. Assuming
$J$ and $J_0$ to be large compared to $|J_a|$, one arrives at
a 'minimal version' of the model, where spins along a
chain have the same sign between two defects, reversing sign
at a defect. The only relevant energy parameter
is $J_a$ \cite{selke,holtschn}.

The model, in its minimal and full variants, is known to
form straight defect stripes, perpendicular to the chains, in
the ground state, $T= 0$, with arbitrary separation
between the stripes. As temperature $T$ is increased
the stripes will meander, tending to keep, on average, their largest
possible distance due to
entropic repulsion. Eventually the stripes will break up
at a phase transition of first order, associated with a pairing
of defects. We introduce a local pinning with strength $E_p$ of
the defects at the sites of straight equidistant lines perpendicular
to the chains. At low temperatures, the stripes stay close
to the pinning lines and
long--range antiferromagnetic order is observed. The transition
remains to be of first order, driven, again, by  the
enhanced pairing of defects \cite{holtschn}.
\begin{figure}
  \includegraphics[width=\linewidth]{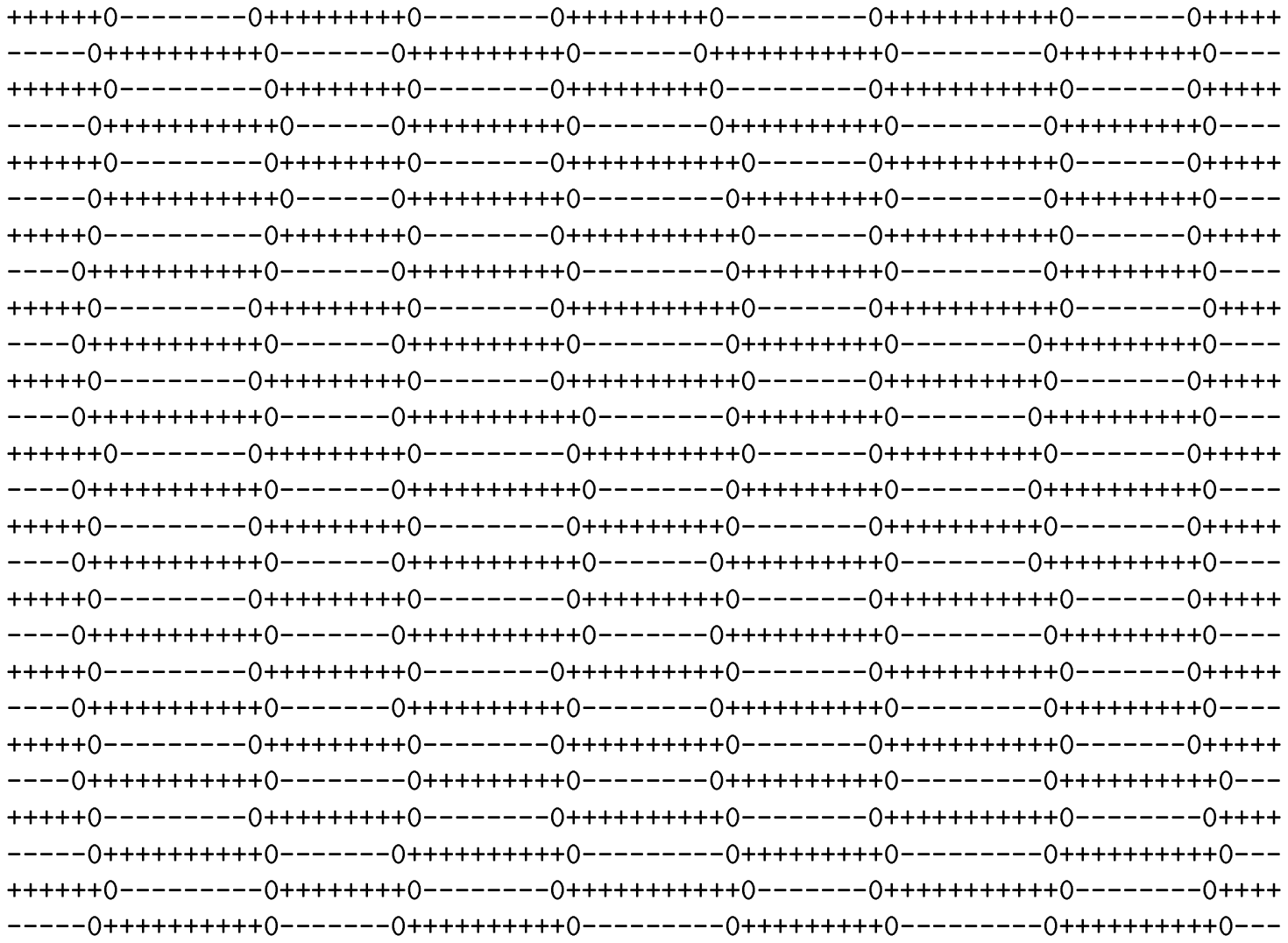}
  \caption{Typical equilibrium configuration with
    zig--zag structures in the minimal Ising
    model, at $k_BT/|J_a|= 0.2$ and
    $E_p/|J_a|=0.1$. Only
    a part of the system with totally $80 \times 80$ sites is shown.}
\end{figure}

Applying now an external field $H$ in the Ising direction, the
ground state continues to consist of straight stripes at
sufficiently low fields. At $H_1 < H < H_2$, assuming
a vanishing or weak pinning potential, the defects
will form zig--zag stripes separating the antiferromagnetic
domains, thereby allowing for a
non--zero total magnetization, see Fig. 1 for
a typical equilibrium configuration at low
temperatures. For $H > H_2$, in the minimal model the
defects will be paired, with isolated
spins between two neighbouring defects pointing opposite
to the field direction. In the full model, even those spins may
be reversed in even stronger fields. Upon increasing the
temperature, at fixed field $H < H_2$, the Ising magnet
will undergo a phase transition at which the defect
stripes, being either straight or of zig--zag type, become
unstable. Indeed,
performing extensive
Monte Carlo simulations, we determined phase diagrams of the
Ising model in the $(T,H)$--plane, as depicted
in Fig. 2 for the minimal variant with different pinning
strengths, $E_p$ \cite{holt2}.
\begin{figure}
  \includegraphics[width=\linewidth]{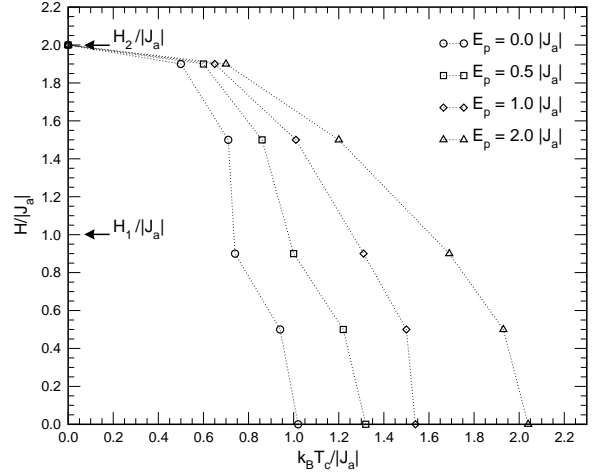}
  \caption{Phase diagram of the minimal Ising model with various
    pinning strengths, $E_p$.}
\end{figure}

In particular, we find no evidence for a transition associated with
the zig--zag structure. Albeit there is a jump in the
total magnetization, and thence a divergence in the susceptibility, at
$T= 0$ and $H= H_1$, these quantities change smoothly even at very low
temperatures. Indeed, small zig--zag segments are thermally excited
well below $H_1$, with a gradual increase of their average
length as $H$ is increased \cite{holt2}.

A large pinning potential may
suppress the zig--zag structures in the minimal and full 
models \cite{holt2}.

We conclude that the
zig--zag structures are presumably not relevant in explaining
the transition from the antiferromagnetic
to the disordered phase in
(La,Ca)$_{14}$Cu$_{24}$O$_{41}$ \cite{amm,klin}. However, a melting of
extended or local defect stripes may play an important role in
that transition.

\section{Anisotropic Heisenberg antiferromagnet}\label{sec3}

Following Matsuda \textit{et al.} \cite{mat}, the magnetic properties
of La$_5$Ca$_9$Cu$_{24}$O$_{41}$ depend
on the Cu$^{2+}$ ions located in the $ac$--planes, having
a centered rectangular geometry, see Fig. 3. Based
on their spin--wave analysis, the spins ($S=1/2$)
of the ions couple along the CuO$_2$ chains, i.e.\ along
the $c$ axis (vertical direction in Fig. 3), through
nearest neighbour, $J_{c1}$, and next-nearest neighbour, $J_{c2}$,
exchange constants, with $J_{c1}=-0.2$ meV being antiferromagnetic
and $J_{c2}=0.18$ meV being ferromagnetic.
The ferromagnetic ordering in the chains is due to the strong
antiferromagnetic interchain couplings: $J_{ac1}=-0.681$ meV refers
to the two nearest neighbours
in the adjacent chain, and  $J_{ac2}=-0.3405$ meV denotes
the couplings to the two next-nearest neighbours.
\begin{figure}
  \includegraphics[width=\linewidth]{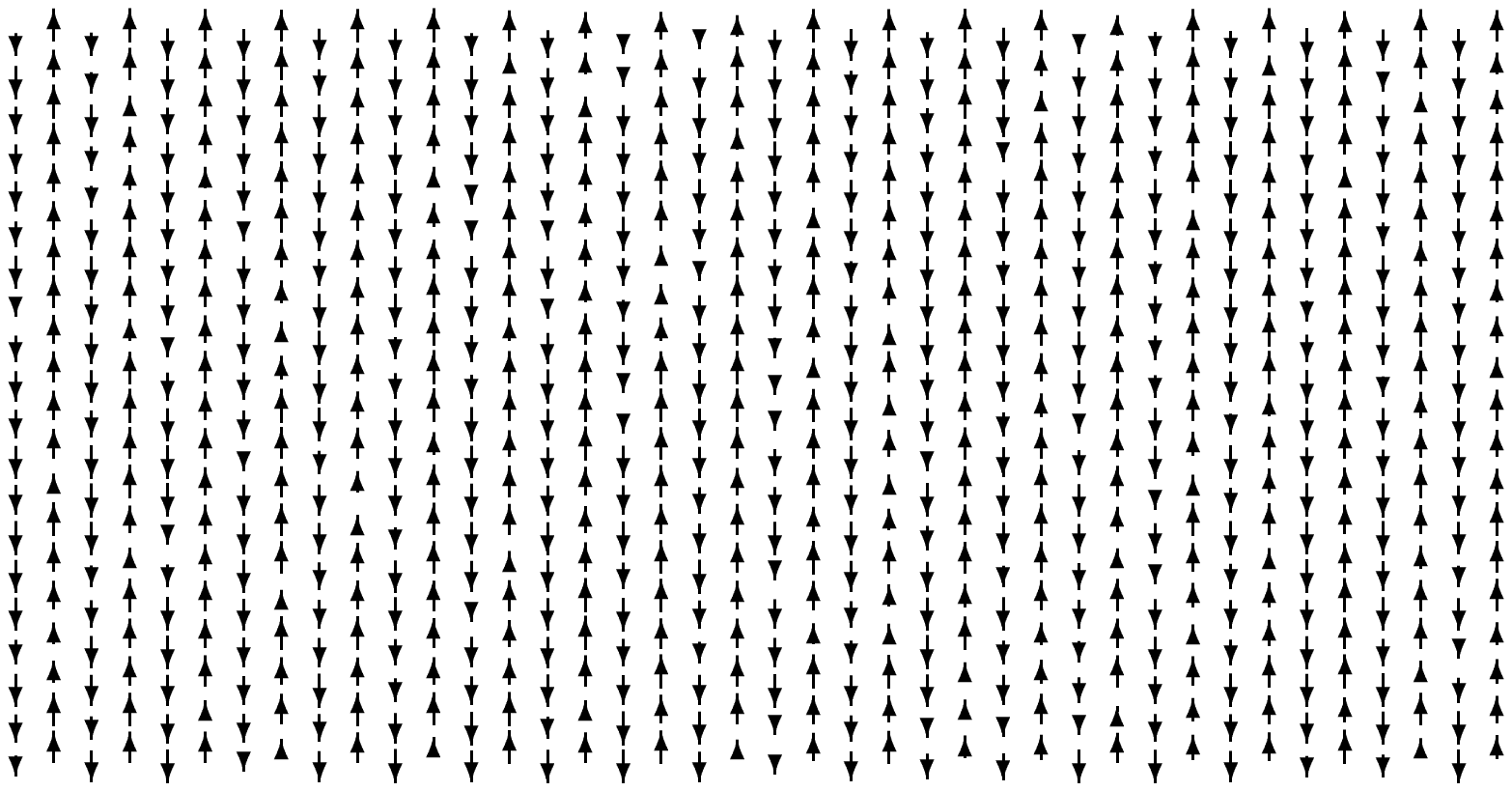}
  \caption{Typical equilibrium configuration of the $z$--com\-po\-nents
    of the spins in the
    antiferromagnetic phase of a classical variant of the anisotropic
    Heisenberg model of Matsuda et. al. \cite{mat} without field, at
    $k_BT$= 0.5 meV. Only a part of the system with totally $40 \times 40$
    sites is shown.}
\end{figure}

Importantly, there is an uniaxial anisotropy favouring alignment of the spins
along the $b$ axis. Its contributions to the different couplings
are not known, and its total  effect may be mimicked in the
classical variant of the model with spins of length one by
a single-ion interaction $D=-0.211$ meV.

We studied the classical model with external
fields along the easy axis, $H_z$, and
perpendicular to it,
$H_x$, doing Monte Carlo simulations. In both cases, one encounters
an antiferromagnetic phase at low fields, see Fig. 3. The complete
phase diagrams, as obtained from finite--size analyses of the
simulational data, are shown in Fig. 4. In the case of $H_x$, the
phase transition is continuous, being in the Ising universality
class. More interestingly, in the case of $H_z$, there is a
spin--flop--phase with algebraically decaying spin correlations. The
boundary of the antiferromagnetic phase is, at low fields, in
the Ising universality
class as well. As suggested by the
behaviour of the Binder cumulant, the transition between the
antiferromagnetic and paramagnetic phases eventually becomes
of first order as the field is increased, with a tricritical
point at about $k_BT \approx 0.79$ meV \cite{leidl2}. At even
stronger fields, one encounters, closeby, at $k_BT \approx$ 0.75
meV \cite{leidl2}, a triple point between these
two and the spin--flop phases.
\begin{figure}
  \includegraphics[width=\linewidth]{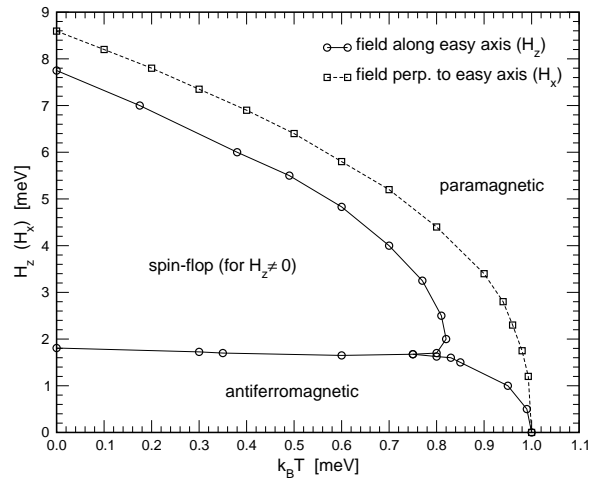}
  \caption{Phase diagrams of the anisotropic Heisenberg model \cite{mat}
    with fields parallel and perpendicular to the easy axis of the spin couplings.}
\end{figure}

Our preliminary Monte Carlo study on the standard classical anisotropic
Heisenberg model with nearest neighbour antiferromagnetic interactions
in two dimensions \cite{bl} suggests that the phase diagram
keeps the same topology for that simpler model, as had been proposed
before for its quantum version \cite{sch}.

Note that experiments on La$_5$Ca$_9$Cu$_{24}$O$_{41}$ do not show a
direct transition between
the antiferromagnetic and spin--flop phases \cite{amm,klin,leidl1}. This
may be due to defects which have not been taken into account 
in the model of Matsuda {\it et al.} Their possible influence
has been discussed elsewhere \cite{leidl2}.\medskip

We thank B. B\"uchner, R. Klingeler, T. Kroll, M. Matsuda, and
V. Pokrovsky for very helpful cooperation and information, and the
Deutsche Forschungsgemeinschaft for financial support.


\begin{thebibliography}{00}
  
\bibitem{miz} Y.\ Mizuno, T.\ Tohyama, S.\ Maekawa, T.\ Osafune,
  N.\ Motoyama, H.\ Eisaki, and S.\ Uchida, Phys.\ Rev.\ B {\bf 57}, 5326
  (1998).
\bibitem{amm} U.\ Ammerahl, B.\ B\"uchner, C.\ Kerpen, R.\ Gross,
  and A.\ Revcolevschi, Phys.\ Rev.\ B {\bf 62}, R3592 (2000).
\bibitem{mat} M.\ Matsuda, K.\ Kakurai, J.\ E.\ Lorenzo, L.\ P.\ Regnault,
  A.\ Hiess, and G.\ Shirane, Phys.\ Rev.\ B {\bf 68}, 060406(R) (2003).
\bibitem{klin} R.\ Klingeler, PhD thesis, RWTH Aachen (2003).
\bibitem{selke} W.\ Selke, V.L.\ Pokrovsky, B.\ B\"{u}chner, and T.\ Kroll, Eur. Phys. J. B {\bf 30}, 83 (2002).
\bibitem{holtschn} M.\ Holtschneider and W.\ Selke, Phys. Rev. E {\bf 68}, 026120-6 (2003).
\bibitem{leidl1} R.\ Leidl and W.\ Selke, Phys. Rev. B {\bf 69}, 056401-2 (2004).
\bibitem{leidl2} R.\ Leidl and W.\ Selke, (submitted to Phys. Rev. B).
\bibitem{bl} D.\ P.\ Landau and K.\ Binder,  Phys.\ Rev.\ B {\bf 24},
  1391 (1981).
\bibitem{holt2} M.\ Holtschneider, Diploma thesis, RWTH Aachen (2004).
\bibitem{sch} G.\ Schmid, S.\ Todo, M.\ Troyer, and A. Dorneich,
  Phys.\ Rev.\ Lett.\ {\bf 88}, 167208 (2002).
\end{thebibliography}
\end{document}